\documentclass[a4paper,twoside]{article}

\usepackage{epsfig}
\usepackage{subfigure}
\usepackage{calc}
\usepackage{amssymb}
\usepackage{amstext}
\usepackage{amsmath}
\usepackage{amsthm}
\usepackage{multicol}
\usepackage{pslatex}
\usepackage{apalike}
\usepackage{SciTePress}
\usepackage[small]{caption}

\begin{document}

\title{\uppercase{CONTROLLER DESIGN FOR A STANCE-CONTROL KNEE-ANKLE-FOOT ORTHOSIS BASED ON OPTIMIZATION TECHNIQUES}  \subtitle{} }

\author{\authorname{S.H. HosseinNia\sup{1}, F. Romero\sup{2}, B. Vinagre\sup{1}, F.J. Alonso\sup{2}, I. Tejado\sup{1}}
\affiliation{\sup{1}Dept. of Electrical, Electronic and Automation Engineering, University of Extremdura, Avda de Elvas S/N, Badajoz, Spain}
\affiliation{\sup{2}Dept. of Mechanical Engineering, Energetics and Materials, University of Extremdura, Avda de Elvas S/N, Badajoz, Spain}
\email{\{hoseinnia;bvinagre;fjas;itejbal\}@unex.es, fromeros@alumnos.unex.es}}

\keywords{Active orthosis, Biomechanics, Muscle modeling, Optimization, Optimal Control}

\abstract{Design of active orthosis is a challenging problem from both the dynamic simulation and control points of view. The redundancy problem of the simultaneous human-orthosis actuation is an interesting exercise to solve concerning the analytical and computational cost effectiveness. The physiological static optimization approach tries to solve the actuation sharing problem. Its objective is to quantify the contributions of muscles and active orthosis to the net joint torques in order to select the proper actuator for the joint. Depending on the disability of each patient, different controllers can be designed. As a matter of fact, the duration of the gait cycle for each patient should be different. In this paper, a PI controller is designed whose parameters are tuned by optimizing a cost function which takes into account the patients muscle power and the error of the knee angle with the reference value. Moreover, the final time is obtained by minimizing the mean of integral squared errors. The performance of the method is shown by designing the controller for three types of patients, ordered from low to high disability. The objective of this work is to use optimal control techniques based on physiological static optimization approach to the design of active orthosis and its control.}

\onecolumn \maketitle \normalsize \vfill

\section{\uppercase{Introduction}}
\label{sec:introduction}

\noindent Spinal cord injuries (SCI) cause paralysis of the lower
extremities because of the break of the connection between nervous central
system and muscular units of the lower body. According to the standard
neurological classification of the American Spinal Injury Association (ASIA),
there are different SCI levels depending on motor and sensory function to be
preserved. The ASIA Impairment SCALE (AIS) range them from A (complete SCI) to
E (normal and sensory function). This work focuses in the assistance of
incomplete SCI subjects with AIS level C or D. Those patients have partially
preserved motor function in the key lower limb muscle groups, and can perform
a low-speed and high-cost pathological gait by using walking aids. The energy
cost and aesthetics of this walk can be performed by means of active orthosis,
requiring external actuation mechanisms to assist the motion of the lower limb
joints during gait cycle. Considerable efforts have been focused on the design
and application of passive and active orthoses to assist standing and walking
of SCI individuals. 

{\normalsize There is a great evolution between the first controllable active,
a patent by Filippi in 1942 \cite{FILIPPI} of a hydraulically-actuated device
for adding power at the hip and knee joints, and the actual orthotic devices.
Concerning the first, developed at the University of Belgrade in the 60's and
70's by Vukobratovich et al. \cite{VUKO}, these early devices to aid people
with paraplegia resulting from spinal cord injury were limited to predefined
motions and had limited success. Nowadays, orthotic systems use predefined
patterns of joint motions and torques together with classical control
techniques or EMG-based control, with the aim of integrating the human
musculoskeletal system and the assisting device. There are different designs
in the literature, see for example the review of Dollar \cite{DOLLAR}.
Nevertheless, few studies \cite{SILVA,KAO} examine the moment joint patterns
of combined patient-orthosis systems. Moreover, the number of studies testing
these systems on handicapped subjects is paradoxically low when comparing with
the studies on able-bodied subjects wearing the orthosis. }

{\normalsize To assist the proper design of active orthoses for incomplete
SCI, it is necessary to quantify the simultaneous contributions of muscles and
active orthosis to the net joint torques of the human-orthosis system.
Simulation of walking in individuals with incomplete SCI wearing an active
orthosis is a challenging problem from both the analytical and the
computational points of view, due to the redundant nature of the simultaneous
actuation of the two systems. In this work, the functional innervated muscles
of SCI patients will be modeled as Hill-type actuators, while the idle muscles
will be represented by stiff and dissipative elements that increment the
passive moments of the inactive joints. The orthosis will be considered as a
set of external torques added to the ankles, knees and hips to obtain net
joint torque patterns similar to those of normal unassisted walking. Kao
\cite{KAO} suggests that able-bodied subjects aim for similar joint moment
patterns when walking with and without robotic assistance rather than similar
kinematic patterns. This is the fundamental hypothesis of this approach to
obtain muscular forces: the combined actuation of the musculoskeletal system
of the SCI subject and the active orthosis produce net joint moment patterns
similar to those of normal unassisted walking. The muscle-orthosis redundant
actuator problem was solved through a physiological static optimization
approach \cite{ALONSO}. A comparison between cost functions and various sets
of innervated muscles can be found in this work. Based on these results, as
Font et al.  explained in \cite{FONT}, the proper actuation can be selected,
but control techniques are required to achieve a suitable gait. }

{\normalsize The objective of this work is to design an optimal controller
based on the minimization of a cost function that takes into account the
patients muscle power and the tracking error of the knee angle. For the
patients with less capability, the weigh of the muscle power in the cost
function will be chosen bigger whereas for patients with more capability, this
weigh will be chosen smaller. Therefore, this controller will consider a trade
off between accuracy of knee movement regarding to healthy human waking and
muscle power of each patient. }

{\normalsize The rest of the paper is organized as follows. In Section 2,
musculoskeletal modeling is stated. In addition, in order to obtain the
muscular power developed by each muscle during gait cycle, the optimization
approach proposed in Alonso et al. \cite{ALONSO} is applied. Section 3
addresses the design of the optimal controller. Finally, Section 4 includes
the main conclusions of this work. }

\section{{\protect\normalsize \uppercase{Musculoskeletal modeling}}}

{\normalsize \noindent In this section, the biomechanical model adopted to
obtain net joint torques for normal walking is presented, as well as the
muscle models for the functional (innervated) and partially denervated muscles
of the spinal cord injured subject.  }

\subsection{{\protect\normalsize Biomechanical multibody model}}

{\normalsize The biomechanical model used has 12 degrees of freedom. It
consists of twelve rigid bodies linked with revolute joints (see Fig.
\ref{fig:Fig1Biomech}), and is constrained to move in the sagittal plane. Each
rigid body is characterized by mass, length, moment of inertia about the
center of mass, and distance from the center of mass to the proximal joint.
The equations of motion of the biomechanical multi-body system can be written
as: }

{\normalsize
\begin{equation}
M\ddot{q}+\Phi_{q}^{T}\lambda=Q,\label{eq1}%
\end{equation}
where $M$ is the global (human-orthosis) mass matrix, $\Phi_{q}^{T}$ is the
Jacobian matrix of the constraint equations, $\ddot{q}$ is the acceleration
vector, $Q$ is the generalized force vector and $\lambda$ are the Lagrange
multipliers. Using kinematic and anthropometric data in (\ref{eq1}), the net
joint torques during a physical activity or motion and the resultant force and
moment due to body-ground contact can be calculated. }

{\normalsize \begin{figure}[ptbh]
{\normalsize
\centering
{\epsfig{file = 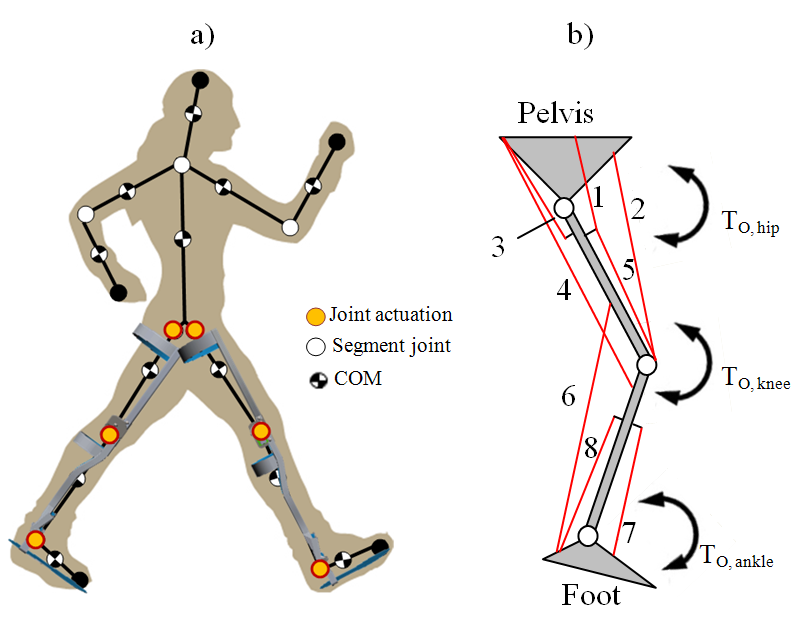, width =0.8 \linewidth}}  }\caption{(a)
Biomechanical model of the human orthosis system. (b) Muscle groups of the
lower limbs: 1 - Iliopsoas, 2 - Rectus Femoris, 3 - Glutei, 4 - Hamstrings, 5
- Vasti, 6 - Gastrocnemius, 7 - Tibialis Anterior, 8 - Soleus.}%
\label{fig:Fig1Biomech}%
\end{figure} }

\subsection{{\protect\normalsize Muscle modeling: innervated and denervated
muscles}}

{\normalsize According to AIS, it is possible to define several levels which
indicate the severity of the injury from A (complete) to E (normal motor and
sensory function). In the design cases C and D, the motor function is
preserved below the neurological level (lowest segment where motor and sensory
functions are normal), being the difference between them the muscle activity
grade of the key muscles. This grade ranges from 0 (total paralysis) to 5
(active movement, full range of motion, normal resistance). As Alonso et al.
proposed in \cite{ALONSO}, the weakness of the denervated muscles is modeled
through a \textit{weakness factor} $p\in\lbrack0,1]$ that limits the maximum
activation of this muscles. }

{\normalsize Both innervated and denervated are modeled as Hill-type
actuators. The Hill-type muscle-tendon model \cite{ZAJAC,WINTER} is shown in
Fig. \ref{fig:MODELO} (a) and \ref{fig:MODELO} (b). It consists of a
contractile element (CE) that generates the force, a nonlinear parallel
elastic element (PE), representing the stiffness of the structures in parallel
with muscle fibers, and a nonlinear series elastic (SE) element that
represents the stiffness of the tendon which is serially attached to the
muscle and completes the muscletendon unit. The two differential equations
that govern the muscle dynamics are: }

{\normalsize
\begin{equation}
\dot{a}=h(u,a), \label{eq2}%
\end{equation}%
\begin{equation}
\dot{f}_{mt}=g(a,f_{mt},l_{mt},v_{mt}). \label{eq3}%
\end{equation}
}

{\normalsize Equation (\ref{eq2}) refers to the activation dynamics, which
relates the muscle excitation $\mathit{u}$ from the central nervous system
(CNS) and the muscle activation $a\in\lbrack0,1]$. On the other hand, equation
(\ref{eq3}) defines the force-generation properties as a function of the
muscle tendon length $l_{mt}$ and velocity $v_{mt}$. Activation dynamics is
not considered for the purpose of this work. }

{\normalsize If the pennation angle $\alpha$ is constant, in accordance with
Fig. \ref{fig:MODELO} (b): }

{\normalsize
\begin{equation}
l_{mt}=l_{se}+l_{ce}cos(\alpha), \label{eq4}%
\end{equation}%
\begin{equation}
f_{mt}=f_{se}=(f_{ce}+f_{pe})cos(\alpha)\approx f_{ce}cos(\alpha), \label{eq5}%
\end{equation}
where the force of the parallel elastic element is set to zero
\cite{ACKERMANN1,ACKERMANN2}. The tendon (SE) can be modeled by a simple
quadratic force-strain curve depending on tendon stiffness as follows:  }

{\normalsize
\begin{equation}
f_{SE}=\left\{
\begin{matrix}
0 & \mbox{if }l_{se}<l_{ts}\\
k_{t}(l_{se}-l_{ts})^{2} & \mbox{if }l_{se}>l_{ts}%
\end{matrix}
\right.  , \label{eq6}%
\end{equation}
where $l_{ts}$ is the tendon slack length and $k_{t}$ is the SE stiffness,
which is given by:  }

{\normalsize
\begin{equation}
k_{t}=\frac{f_{0}}{(\epsilon_{0}l_{t}s)^{2}}, \label{eq7}%
\end{equation}
being $\epsilon_{0}$ ($3$\% to $5$\%) the strain occurring at the maximal
isometric muscle force $f_{0}$ \cite{ACKERMANN1}. The force generated by the
CE $f_{ce}$ is a function of the activation $a$, its length $l_{ce}$, and its
contraction velocity $v_{ce}$. The expression for the concentric contraction
($v_{ce}<0$) reads as:  }

{\normalsize
\begin{equation}
\frac{f_{ce}}{f_{0}}=a\frac{B_{r}(f_{iso}+A_{r})-A_{r}(B_{r}-\frac{\tilde
{v}_{ce}^{N}}{f_{ac}})}{B_{r}-\frac{\tilde{v}_{ce}^{N}}{f_{ac}}}, \label{eq8}%
\end{equation}
where $\tilde{v}_{ce}^{N}=v_{ce}/l_{ce}^{opt}$, $A_{r}=0.41$, $B_{r}=0.52$ and
$f_{iso}=f_{iso}(w,l_{ce}^{opt},l_{ce})$, which corresponds to the muscle
isometric force relative to the maximal isometric muscle force $f_{0}$ and
$f_{ac}=min(1,3.33a)$.  }

{\normalsize The expression for the eccentric contraction ($v_{ce}>0$) depends
on $\tilde{v}_{ce}^{N}$ and $f_{iso}$. The force-length-velocity relationship
is shown in Fig. \ref{fig:MODELO} (c). }

{\normalsize \begin{figure}[h]
{\normalsize \centering
{\epsfig{file = 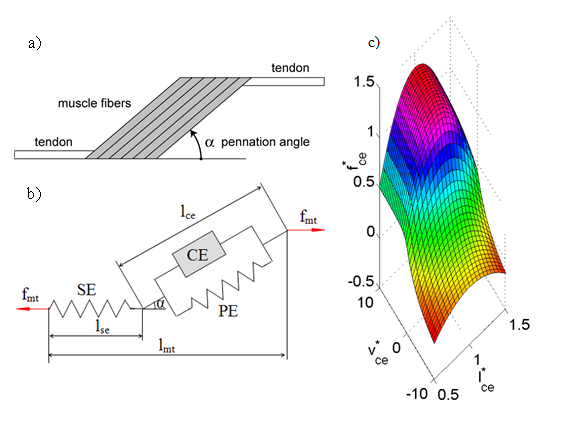, width = 0.9 \linewidth}}  }\caption{Muscle model:
(a) Conceptual scheme. (b) Hill model \cite{ZAJAC}. (c) Force-length-velocity
model.}%
\label{fig:MODELO}%
\end{figure} }

{\normalsize In order to quantify the muscle weakness, the muscle activation
will be multiplied by the mentioned weakness factor $p$, where $p=1$ for
innervated muscles, $0<p<1$ for partially denervated muscles and $p=0$ for
totally denervated muscles (no activity). The atrophy of denervated muscles,
as exposed by Thomas and Grumbles \cite{THOMAS}, depends on the elapsed time
from the injury. This atrophy increases the passive torques at the joint.
Several studies \cite{EDRICH,LEBIEDOWSKA,AMANKWAH} show that passive torques
tend to be larger in pathological than in healthy individuals. To take this
fact into account stiff and dissipative elements are included into the model
using the definitions given in \cite{AMANKWAH}. Fig. \ref{fig:KNEE} shows the
increment of the knee torque due to pathological passive torque compared with
the torques in normal gait (obtained through inverse dynamics analysis from
the 2D walking kinematic benchmark from Winter \cite{WINTER}). As can be
observed, only very slight differences can be found between both torques. }

{\normalsize \begin{figure}[h]
{\normalsize \centering
{\epsfig{file = 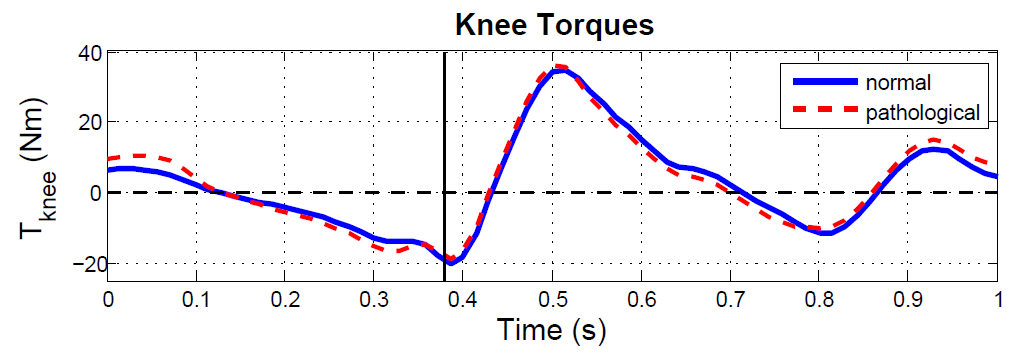, width =  \linewidth}}  }\caption{Knee torques during
normal gait (solid blue line) and pathological gait (dashed red line) for the
right leg. Swing phase occurs from 0 to 0.4 in normalized time and stance
phase from 0.4 to 1.}%
\label{fig:KNEE}%
\end{figure} }

\subsection{{\protect\normalsize Optimization approach. Muscle and orthosis
actuation}}

{\normalsize In order to solve the load sharing problem in biomechanics,
optimization procedures are used next. There are several optimization methods
(static and dynamic optimization, large-scale optimization) and optimization
criteria (minimization of the metabolic cost of transport, minimization of
muscle stresses) in the literature \cite{MENEGALDO,YAMAGUCHI}. In order to
obtain the forces that will be used in the design of the controller, we use
the physiological static optimization approach \cite{ALONSO}. }

{\normalsize This modified version of the classical static optimization
approach considers muscle contraction dynamics, ensuring the physiological
consistency of the solution. This approach comprises two steps. In the first
one, the inversion of the contraction dynamics is solved assuming that muscle
activation are maxima. The length ($l_{mt}$) and velocity ($v_{mt}$) of each
tendon unit involved in the process are obtained from the generalized
coordinates of the multi-body model and the maximum muscle-tendon length
\cite{GERRITSEN}. Then, the maximum muscle force histories $f_{mt}^{\ast}(t)$
compatible with contraction dynamics are calculated assuming the muscle
activation is maxima at each instant, i.e. $A_{m}=[a_{1},\ldots,a_{N}%
]^{T}=[1,\ldots,1]^{T}$. Briefly, for each muscle, the contraction dynamics
differential equation is integrated as:  }

{\normalsize
\begin{equation}
\frac{df_{mt}^{\ast}}{dt}=g((a=1)\cdot p,f_{mt}^{\ast},l_{mt},v_{mt}).
\label{eq9}%
\end{equation}
}

{\normalsize In the second step, the muscle activations and orthosis actuation
is calculated by solving the optimization problem: {\small
\begin{equation}%
\begin{matrix}
\label{eq10}Min & J(F_{mt},T_{o})=\omega_{mt}\sum_{j=1}^{8}{(-f_{ce,j}%
v_{ce,j})^{2}}+\omega_{0}\sum_{k=1}^{3}{(T_{o,k}\dot{\theta}_{k})^{2}}\\
s.t. & R\cdot(AF^{\ast})=T\\
& 0\leq a_{j}\leq1,j=1,\ldots,N=8\\
& -1\leq o_{k}\leq1,k=1,2,3
\end{matrix}
,
\end{equation}
}where $AF_{mt}^{\ast}=[a_{1}\cdot p_{1}\cdot f_{mt,1}^{\ast},\ldots
,a_{8}\cdot p_{8}\cdot f_{mt,8}^{\ast},o_{1}\cdot T_{o,1}^{\ast},\ldots
,o_{3}\cdot T_{o,3}^{\ast}]$.  }

{\normalsize With this approach, muscle forces and orthosis actuation are
calculated for a gait cycle in order to optimize the cost function and obtain
the parameters of the PI controller proposed. The 2D walking kinematic
benchmark data from Winter \cite{WINTER} was used to perform an inverse
dynamic analysis. This movement corresponds to a healthy female subject with
57.75 kg of weight with normal gait. Once joint torques have been calculated,
the optimization problem is solved by using MATLAB routine \textit{fmincon}
implemented in the optimization toolbox that uses a Sequential Quadratic
Programming (SQP) method. The simulated muscle-orthosis actuation was
performed for an AIS C subject: motor function partially preserved below the
neurological level and more than half of the key muscles below the
neurological level have a muscle grade less than $3$. To simulate this kind of
injury, we have defined the following vector of weakness factor:
\[
p=[1,0.2,1,0.2,0.2,0.2,0.2,0.2,]^{T}.
\]
}

{\normalsize The orthosis actuation prevents stance phase knee flexion due to
quadriceps and assists swing-phase flexion depending on the ability of each
patient, as shown in the simulation results for the knee in Fig. \ref{par}. }

\section{{\protect\normalsize Controller design}}

{\normalsize In order to control the orthosis, a mathematical model of the motor is needed.
In particular, the following second order transfer function is used (see
\cite{HosseinNia_11}):
\begin{equation}
P\left(  s\right)  =\frac{\theta_{o}}{V_{in}}=\frac{3.58}{s\left(
0.01s+1\right)  }.\label{sys}%
\end{equation}
}

{\normalsize To control the orthosis, a classic PI controller is considered
as:
\begin{equation}
C(s)=k_{p}+\frac{k_{i}}{s}. \label{PI_C}%
\end{equation}
}

{\normalsize The aim is to tune the parameters $k_{p}$ and $k_{i}$ in order to
optimize the following cost function,
\begin{equation}
J=\int{\beta(-f_{mt}v_{mt})^{2}+(1-\beta)e^{2},}\label{Costf}%
\end{equation}
with $e=\theta_{k_{ref}}-\theta_{k}$. This cost function consists of two
parts. The first one corresponds to the muscle power where $f_{mt}$ is the
muscular forces obtained by optimization and $v_{mt}$ the muscular velocities.
The second part refers to the error between the knee angle $\theta_{k}$ and
the reference knee angle $\theta_{k_{ref}}$, respectively. Two weights
$0<\beta<1$ and $(1-\beta)$ can be chosen regarding to the muscle power of the
patients. The idea is to design an optimal controller based on the patients
muscle power and the tracking of a reference signal, where, for the patients
with less muscular capacity, $\beta$ will be chosen bigger in order to
minimize power and perform the movement and, for patients with more capacity,
the value of $\beta$ will be minor, so the cost function prioritizes the
minimization of the tracking error and the movement is going to be made in
less time. Therefore, this controller will consider a trade off between
accuracy of knee movement (concerning healthy human walking) and muscle power
of each patient, taking into account that patients with less capability need
more time to perform the same movement. }

{\normalsize In order to show the performance of the proposed method, the
controller will be designed regarding to the following three weight options:
}

\begin{itemize}
\item {\normalsize $\beta=0.1$ for the patients with low disability,  }

\item {\normalsize $\beta=0.5$ for the patients with fair disability,  }

\item {\normalsize $\beta=0.9$ for the patients with high disability.  }
\end{itemize}

{\normalsize {\small \begin{table}[ptb]
\caption{The optimized controller parameters}%
\label{PIparam}
\begin{center}
{\normalsize {\small
\begin{tabular}
[c]{|l|c|c|}\hline
$\beta$ & $k_{p}$ & $k_{i}$\\\hline
$0.1$ & $40$ & $45$\\\hline
$0.5$ & $15$ & $20$\\\hline
$0.9$ & $1$ & $6$\\\hline
\end{tabular}
}}
\end{center}
\end{table}} }

{\normalsize \begin{figure}[ptbh]
\begin{center}
{\normalsize \includegraphics[width=\linewidth]{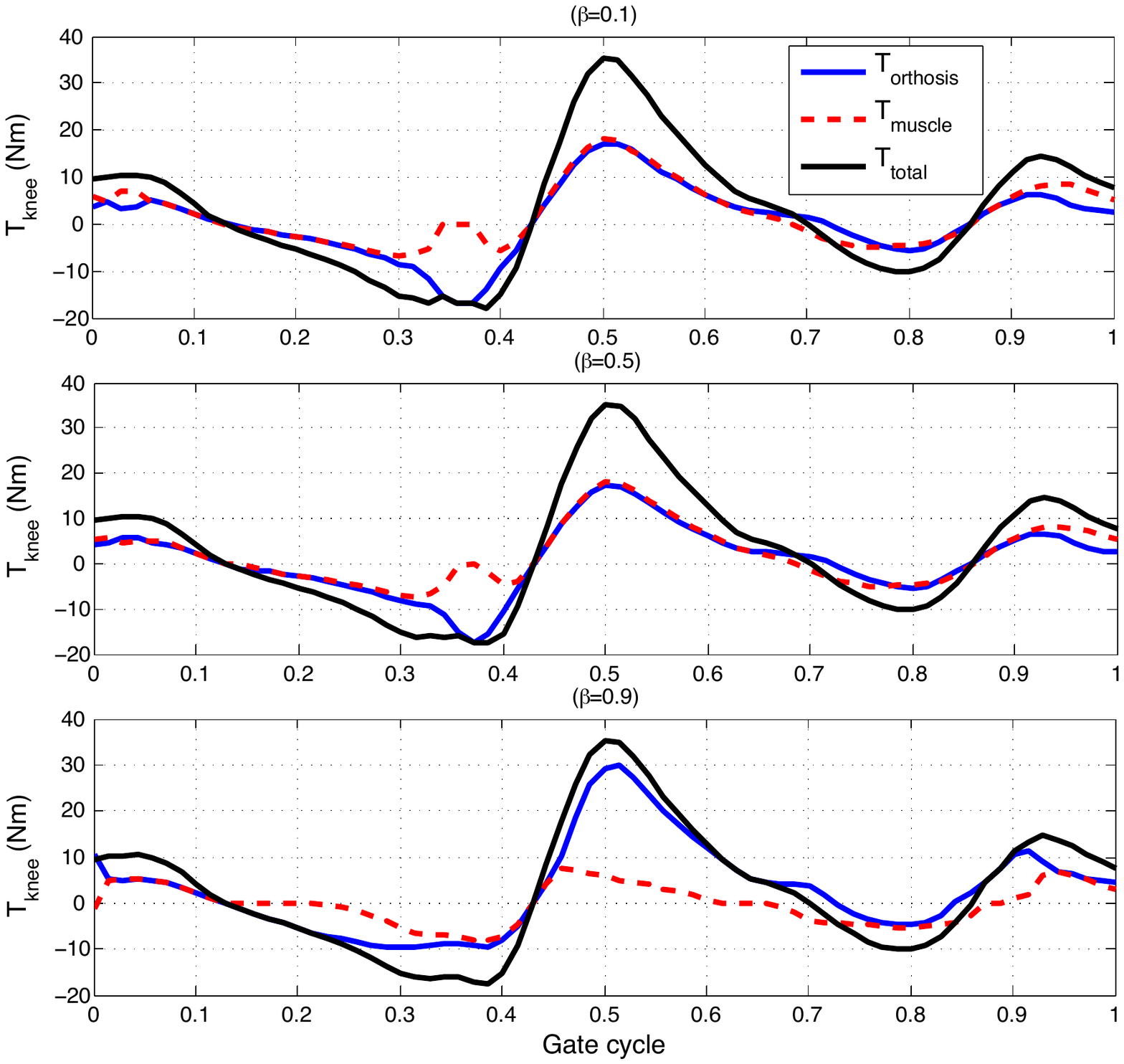}  }
\end{center}
\par
{\normalsize  }\caption{Comparsion of orthosis and muscle torque according to
the patients ability}%
\label{par}%
\end{figure} }

{\normalsize Fig. \ref{par} shows the orthosis torque, muscle torque and total
torque corresponding to each controlled system (the controller parameters ara
given in Table \ref{PIparam}). As can be seen, prioritizing the muscular power
with a big value of $\beta$, we consider that patients have a major disability
and the assistance torque provided by the orthosis should be higher to
compensate the deficiency. On the other hand, prioritizing the tracking error
with a low value of $\beta$ means that patients have more ability to perform
the movement, so the torque provided is lower and the movement is achieved in
less time with more accuracy. The final times corresponding to each case are
calculated based on minimizing the following mean of integral squared errors
(}${\normalsize MISE}${\normalsize ): }

{\normalsize
\begin{equation}
MISE=E\left\Vert \theta_{k}-\theta_{k_{ref}}\right\Vert _{2}^{2}=E\int
_{0}^{t_{f}}\left(  \theta_{k}(t)-\theta_{k_{ref}}(t)\right)  ^{2}%
dt,\label{MISE}%
\end{equation}
where $t_{f}$ denotes the final time in a gate cycle and $E$ denotes the
expected value with respect to that sample. Minimizing }${\normalsize MISE}%
${\normalsize  for each controller designed, the final time is obtained based
on an optimization program to satisfy $MISE<0.01$): \newline%
{\small \begin{table}[ptb]
\caption{Final time}%
\label{PIparam}
\begin{center}
{\normalsize {\small
\begin{tabular}
[c]{|l|c|c|}\hline
$\beta$ & Final time & \\\hline
0.1 & 0.75 & \\\hline
0.5 & 1.4 & \\\hline
0.9 & 6.1 & \\\hline
\end{tabular}
}}
\end{center}
\end{table}} }

{\normalsize \begin{figure}[ptbh]
\begin{center}
{\normalsize \includegraphics[width=\linewidth]{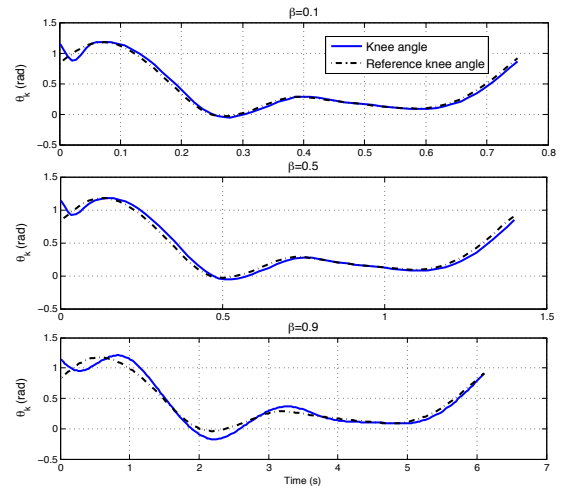}  }
\end{center}
\par
{\normalsize  }\caption{Effect of disability in final time}%
\label{kneetf}%
\end{figure} }

{\normalsize Fig. \ref{kneetf} shows the effect of the disability of the
patients in final times. As can be observed, patients with higher $\beta$
(considered with disability) need more time to perform the complete gait cycle
than patients with low value of $\beta$, corresponding to low disability. In
the same way, high disability correspond to higher tracking error, and lower
disability correspond to a better accuracy in the performance of the movement
compared with the healthy subject. }

\section{{\protect\normalsize \uppercase{Conclusions}}}

{\normalsize \label{sec:conclusion} \noindent In this paper, in order to
control an orthosis, an optimal approach is proposed to design a PI controller
according to disability of the patients. This disability is simulated by means
of physiological static optimization approach where the muscular forces of SCI
are obtained in a process that combines the actuation of the muscles and the
external actuation provided by the orthosis. Those forces are used to design a
proper controller for the external actuation. Considering patients with a high
disability, the controller is tuned to perform the movement so as to allow the
patient to achieve the movement but in a longer cycle compared with patients
with less disability, where the controller is tuned giving priority to the
accuracy of the movement. Patients with less power in his muscle --high
disability--, need more time in a gate cycle to walk, whereas patients with
low disability need less time. This idea is shown through some three types of
disability, i.e. high disability, fair disability and low disability. The
simulation results show the efficiency of the proposed method. }

\section*{{\protect\normalsize \uppercase{Acknowledgements}}}

{\normalsize \noindent This work was supported by the Spanish Ministry of
Science and Innovation under the project DPI2009-13438-C03. The support is
gratefully acknowledged. }

\bibliographystyle{apalike}
\bibliography{BIODEVICES2012}

\end{document}